\documentclass[aps,prc,amsmath,amssymb,preprintnumbers,nofootinbib,twocolumn,showpacs,floatfix,10pt]{revtex4}
\usepackage{epsfig,graphics,color}
\usepackage{graphicx}% Include figure files
\usepackage{dcolumn}% Align table columns on decimal point
\usepackage{bm}% bold math
\usepackage{amsmath}

\newcommand \tie {{\it i.e.}}

\newcommand \ra  {\rightarrow}

\newcommand \h {\theta}

\newcommand \vecr {\vec{r}}

\newcommand \e {\epsilon}

\newcommand \A {\alpha}

\newcommand \D {\Delta}
\newcommand \sg {\sigma}

\newcommand \nt {\noindent}

\newcommand \bvec{\left( \begin{array}{c} }
\newcommand \evec{\end{array} \right)}

\newcommand \bea{\begin{eqnarray} }
\newcommand \eea{\end{eqnarray} }
\newcommand \nn {\nonumber}
\newcommand {\be} {\begin{equation}}
\newcommand {\ee} {\end{equation}}

\newcommand {\mbx} {\mbox{}}

\newcommand {\ata} {& \times &}

\begin{document}

\title{Resolving the plasma profile via differential single inclusive suppression} 

\author{A.~Majumder}
\affiliation{Department of Physics, Duke University, Durham, NC 27708}

\date{ \today}

\begin{abstract}
The ability of experimental signatures to resolve the spatio-temporal profile of an 
expanding quark gluon plasma is studied. In particular, the single inclusive suppression of high 
momentum hadrons versus the centrality of  a heavy-ion collision and with respect to 
the reaction plane in non-central collisions is critically examined. Calculations are 
performed in the higher twist formalism for the modification of the fragmentation 
functions. Radically different nuclear geometries are used. The influence of different  initial gluon distributions 
as well as different temporal evolution scenarios on the single inclusive suppression of high momentum 
pions are outlined. It is demonstrated that the modification versus the reaction plane 
is quite sensitive to the initial spatial density. Such sensitivity remains even in the presence of 
a strong elliptic flow. 
\end{abstract}


\maketitle

%%%%%%%%%%%%%%%%%%%%%%%%%%%%%%%%%%%%%%%
%%%%%%%%%%%%%%%%%%%%%%%%%%%%%%%%%%%%%%%
%%%%%%%%%%%%%%%%%%%%%%%%%%%%%%%%%%%%%%%
%%%%%%%%%%%%%%%%%%%%%%%%%%%%%%%%%%%%%%%
%%%%%%%%%%%%%%%%%%%%%%%%%%%%%%%%%%%%%%%

% \section{Introduction}

%%%%%%%%%%%%%%%%%%%%%%%%%%%%%%%%%%%%%%%
%%%%%%%%%%%%%%%%%%%%%%%%%%%%%%%%%%%%%%%
%%%%%%%%%%%%%%%%%%%%%%%%%%%%%%%%%%%%%%%
%%%%%%%%%%%%%%%%%%%%%%%%%%%%%%%%%%%%%%%
%%%%%%%%%%%%%%%%%%%%%%%%%%%%%%%%%%%%%%%

The goal of ultra-high-energy heavy-ion collisions is the creation and study of 
strongly interacting matter, heated past a temperature beyond which confinement
can no longer be expected~\cite{Karsch:2003jg}. 
The Relativistic Heavy Ion Collider (RHIC) at Brookhaven National Laboratory (BNL), 
has provided a wealth of data which call for detailed theoretical 
modeling and study of the produced matter~\cite{RHIC_Whitepapers}. 
At present, computations of the wide variety of  bulk observables~\cite{Braun-Munzinger:2003zd, Kolb:2003dz,Huovinen:2003fa} 
have led to a picture that the Quark-Gluon Plasma (QGP) formed at 
RHIC assumes the dynamical properties of an ideal fluid~\cite{Teaney:2003kp}. 
The study of hard probes~\cite{Gyulassy:2003mc} complements 
this view  by ascribing considerable opaqueness to the produced matter towards the 
passage of hard jets through it. 
This set of observables, when taken in its entirety, has led to the suggestion 
that the matter formed at RHIC is strongly interacting~\cite{Gyulassy:2004zy}.

The object of this letter is to study the feasibility of correlating the 
properties of the plasma derived from investigations in the bulk sector 
with those gleaned from the hard probe sector.  
Full 3+1 dimensional hydrodynamic simulations \cite{Nonaka:2006yn} have performed very well in 
comparison to experimental data in the soft sector. In essence, these simulations make 
predictions regarding the space-time profile (STP) of the expanding matter, but require, 
as input, an initial spatial density profile and are 
sensitive to the ansatz used. 
The possibility of testing such a space-time profile via an independent set of 
measurements afforded by  the modification of the fragmentation of hard jets 
as they pass through the dense matter forms the focus of  the current study. 
Such a study is similar, in spirit, to Ref.~\cite{Armesto:2004pt} which explored the 
effect of the longitudinal flow on the modification of 
hard jets. This complementary effort  will assume boost invariance 
and will be restricted to mid-rapidity.

There exists a wide variety of  observables involving particles at high transverse momentum $p_T$ 
which may offer insight 
regarding the space-time profile of the matter produced in a heavy-ion collision: 
single inclusive observables (both differential and 
integrated)~\cite{highpt,Gyulassy:2000fs}, double inclusive hadronic observables (both near  and away side)~\cite{Adams:2006yt,Vitev:2005yg}, as well as 
photon hadron correlations~\cite{Renk:2006qg}.  In this letter, 
the ability of single inclusive observables to reveal the space-time profile of the plasma will be critically examined. 
It has already been pointed out that the nuclear modification factor $R_{AA}$ as a 
function of $p_T$ for the most central event is not very sensitive to the detailed nature of the 
initial gluon density or the time evolution of this density~\cite{Renk:2006qg}. The induction 
of such sensitivity requires the use of more differential probes.
In this letter,  the nuclear modification versus the reaction plane in non-central 
collisions will be examined. 
A simplified methodology will be followed: While both the production cross section 
of a jet  (which depends on the number of binary collisions $N_{\textrm{bin}}$) 
as well as the density of the plasma at a certain 
transverse location (which is more closely related to the number of participants $N_{\textrm{part}}$) 
depend on the nuclear density profile, these will be disassociated from one another. 
The nuclear density profile relevant to the estimation of the number of initial binary collisions may 
be determined independently in $pA$ collisions. For the current treatment, this will be approximated 
by a hard sphere distribution and left unchanged. 
As the focus is on the dependence of the jet modification on the profile of the produced matter in the 
final state, different nuclear densities and temporal dependences will be used as inputs to produce a space-time 
modulation of the produced matter. 
In this way, a direct measure of the density profile-dependent modification on an 
identical set of produced jets will be explored.   

%%%%%%%%%%%%%%%%%%%%%%%%%%%%%%%%%%%%%%%
%%%%%%%%%%%%%%%%%%%%%%%%%%%%%%%%%%%%%%%
%%%%%%%%%%%%%%%%%%%%%%%%%%%%%%%%%%%%%%%
%%%%%%%%%%%%%%%%%%%%%%%%%%%%%%%%%%%%%%%
%%%%%%%%%%%%%%%%%%%%%%%%%%%%%%%%%%%%%%%

% \section{Formalism}

%%%%%%%%%%%%%%%%%%%%%%%%%%%%%%%%%%%%%%%
%%%%%%%%%%%%%%%%%%%%%%%%%%%%%%%%%%%%%%%
%%%%%%%%%%%%%%%%%%%%%%%%%%%%%%%%%%%%%%%
%%%%%%%%%%%%%%%%%%%%%%%%%%%%%%%%%%%%%%%
%%%%%%%%%%%%%%%%%%%%%%%%%%%%%%%%%%%%%%%

The phenomenology of particle production at very high energies in 
$pp$ collisions is greatly simplified by the factorization theorems of 
QCD \cite{Collins:1989gx}. These predict that at high transverse momentum $p_T$ 
(and as a result large virtuality $Q^2$)  the single inclusive cross-section achieves 
a factorized form. 
 For  high enough transverse momentum $p_T > 7$GeV, further simplifications arise in the case of 
nuclear collisions, \tie, one may ignore a variety of initial and final state nuclear effects, such as recombination, 
 intrinsic transverse momentum and the Cronin effect.  
At such high energies these are known to be small 
compared to the dominant effect of medium induced energy loss which leads to the modification of the 
fragmentation function~\cite{Wang:2003aw}. Such effects may have an interesting density 
dependence and will reappear in any effort to extend the effects discussed herein to lower $p_T$. 
 Assuming a factorization of initial and final state effects, the differential cross-section 
for the production of a high $p_T$ hadron at midrapidity from the impact of 
two nuclei $A$ and $B$ at an impact parameter between $b_{min},b_{max}$ is given as, 

\bea
\frac{d \sigma^{AB}}{dy d^2 p_T} 
 &=& K \int_{\mbx_{b_{min}}}^{\mbx^{b_{max}}}\!\!\!\!\!\!\!\!\!d^2 b 
\int d^2 r t_A(\vec{r}+\vec{b}/2) t_B(\vec{r} - \vec{b}/2)  \nn \\
&\times& \!\!\!\! \int \!\!\! d x_a d x_b  G^A_a(x_a,Q^2)  G^B_b(x_b,Q^2)  \nn \\
\ata \!\!\!\! \frac{  d \hat{\sg}_{ab \ra cd} }{ d \hat{t}}\frac{ \tilde{D}_c^h(z,Q^2)}{\pi z},  \label{AA_sigma}
\eea
\nt
where, $G^A_a(x_a,Q^2)  G^B_b(x_b,Q^2)$ represent the nuclear parton distribution functions. 
These are given in terms of the shadowing functions $S_A(x_a,Q^2)$ 
\cite{Li:2001xa} and the parton distribution functions in a nucleon $G_a(x_a,Q^2)$ as 

\bea
G_a^A(x_a,Q^2) = S_A(x_a,Q^2) G_a(x_a,Q^2). \label{shadow}
\eea
\nt 
In Eq.~\eqref{AA_sigma}, $t_A(\vec{r}),t_B(\vec{r})$ 
are the thickness functions of nuclei $A$ and $B$ at the transverse location $\vec{r}\equiv(x,y)$. 
As mentioned previously, throughout this letter, such thickness functions will be evaluated 
using the simple and analytically 
tractable hard sphere densities ($\rho(x,y,z)=\rho_0 \h(R_A^2 - x^2 - y^2 - z^2)$). 
In this formulation, $\hat{s},\hat{t},\hat{u}$ refer to the Mandelstam 
variables of the internal  partonic process. Unless, stated otherwise, the variables without the 
hats refer to the variables of the full process. The two participating partons in the initial 
state are referred to as $a,b$ while the final state partons are referred to as 
$c,d$. The factor $K\sim 2$ accounts for  higher order contributions.

The most important element in Eq.~\eqref{AA_sigma} is the medium modified 
fragmentation function $\tilde{D}(z,Q^2)$ expressed as the sum of the leading 
twist vacuum fragmentation 
function and a correction brought about by rescattering of the struck 
quark  in the medium \tie, $ \tilde{D} = D + \D D$ \cite{guowang}. 
The vacuum fragmentation functions $D(z,Q^2)$ are taken from 
Ref.~\cite{bin95}. 
In the collinear limit, the modification is computed by isolating corrections, suppressed by powers of $Q^2$, 
which are enhanced by the length of the medium\cite{lqs}. At next-to-leading twist, the correction for the 
fragmentation of a quark has the expression (generalized from deep-inelastic scattering (DIS) \cite{guowang,maj04e}), 

\begin{eqnarray}
\D D(z,Q^2 ,\vecr) &=& \frac{\A_s}{2\pi} \int \frac{dl_\perp^2}{l_\perp^2} 
\int \frac{dy}{y} P_{q \ra i}(y)  2\pi \A_s C_A           \nn \\
\ata T^{M}(\vec{b},\vecr,x_a,x_b,y,l_\perp)D_i \left( \frac{z}{y} , Q^2 \right) \nn \\ 
\ata \left[\frac{\mbox{\Large}}{\mbox{\Large}} \right. l_\perp^2 N_c 
 t_A(\vec{r}+\vec{b}/2) t_B(\vec{r} - \vec{b}/2) \nn \\ 
 \ata \left.  G^A_a(x_a) G^B_b(x_b) 
\frac{d \hat{\sg}}{d \hat{t}} \right] ^{-1}  +  v.c. 
\end{eqnarray}

\nt
In the above equation, $l_\perp$ is the transverse momentum of the radiated gluon (quark) which 
leaves a momentum fraction $y$ in the quark (gluon) denoted as parton $i$ ($P_{q\ra i}(y)$ is the splitting function for 
this process) which then fragments leading to 
the detected hadron. The $v.c.$ refers to  
virtual corrections. 
The factor $T^{M}$ originates from the higher twist matrix element, which encodes the 
information of  rescattering off the soft gluon fields, 

\begin{eqnarray}
T^{M} &=&   t_A(\vecr + \vec{b}/2)  t_B(\vecr -\vec{b}/2)
G^A_a(x_a) G^B_b(x_b)  \frac{d \hat{\sg}}{d \hat{t}}   \nn \\
 \ata \int_0^{\zeta_{max}} \!\!\!\!\!\! d \zeta x_g \rho_g(x_g,\hat{n}\zeta + \vecr) (2 - 2 \cos(\eta_L \zeta))  .
\end{eqnarray}

\nt
In the above equation, the factor 
$\eta_L = l_\perp^2/ [2 \hat{p}_T y(1-y)]$. 
Where, $\hat{p}_T$ represents the transverse 
momentum of the produced parent jet. 
It should be pointed out that in the higher twist formalism used here, no assumption 
is made regarding the prevalent degrees of freedom of the produced matter. 

Breaking with usual practice, the jet direction ($\hat{n}$) is chosen as the 
$x$-axis in the transverse $(x,y)$ plane (the $z$-direction is set by the beam line); 
the angle of the reaction plane 
vector $\vec{b}$ is measured with respect to this direction.  The distance travelled 
by the jet in this direction prior to 
scattering off a gluon is denoted as $\zeta$.  The gluon's forward momentum 
fraction in denoted as $x_g$. 
One assumes, following Ref.~\cite{Baier:1996sk} that the relevant values of $x_g$ are small enough that $x_g\rho_g(x_g, \hat{n}\zeta) $
 is almost independent of $x_g$ and the combination will henceforth be referred to as the density $\rho$. 
This represents 
the one unknown in the above set of equations and in the higher twist formalism. It is 
related to the more familiar quantity $\hat{q}$ by the simple equation \cite{Baier:1996sk}, 

\begin{equation}
\hat{q}(\vecr + \hat{n}\zeta) =  \frac{4 \pi^2 \alpha_s C_F}{N_c^2 - 1} \rho(x_g, \vecr + \hat{n}\zeta) .
\end{equation}

\nt
This is the $\hat{q}$ at a location (and time) $\vecr + \hat{n}\zeta$ in the plasma, and not the averaged value. 

Ignoring the modification in the fragmentation function ($\D D$) and the factors of nuclear 
shadowing in Eq.~\eqref{AA_sigma}, produce the result of binary scaling the $p-p$ scattering 
cross section and produce the denominator of the nuclear modification factor, 

\begin{eqnarray}
R_{AA} &=& \frac{\frac{d \sigma^{AA}}{dy d^2 p_T} }
{T_{AA} (b_{min},b_{max})\frac{d \sigma^{pp}(p_T,y)}{dy d^2 p_T}}. \label{raa}
\end{eqnarray}

The phenomenological input that is required to understand the variation of the nuclear modification 
factor with centrality or $p_T$ is the space time dependence of the gluon density $\rho(x,y,z,\tau)$. 
In the remainder, the focus will be on observables at midrapidity and thus the $z$-coordinate 
will be ignored; different forms of the gluon density profile will be 
invoked and their effect on the nuclear modification factor versus centrality and the reaction 
plane will be elucidated.

%%%%%%%%%%%%%%%%%%%%%%%%%%%%%%%%%%%%%%%
%%%%%%%%%%%%%%%%%%%%%%%%%%%%%%%%%%%%%%%
%%%%%%%%%%%%%%%%%%%%%%%%%%%%%%%%%%%%%%%
%%%%%%%%%%%%%%%%%%%%%%%%%%%%%%%%%%%%%%%
%%%%%%%%%%%%%%%%%%%%%%%%%%%%%%%%%%%%%%%

% \section{The plasma space time profile and the $R_{AA}$}

%%%%%%%%%%%%%%%%%%%%%%%%%%%%%%%%%%%%%%%
%%%%%%%%%%%%%%%%%%%%%%%%%%%%%%%%%%%%%%%
%%%%%%%%%%%%%%%%%%%%%%%%%%%%%%%%%%%%%%%
%%%%%%%%%%%%%%%%%%%%%%%%%%%%%%%%%%%%%%%
%%%%%%%%%%%%%%%%%%%%%%%%%%%%%%%%%%%%%%%

The initial spatial profile of the plasma density depends on the 
nucleon density brought in by the colliding nuclei. It, most likely, should be a 
smooth, monotonously dropping function of the transverse coordinates $(x,y)$,
the time $\tau$ as well as the impact parameter of the collision $b$. The 
function must approach zero as any of these parameters are increased from 
their minimum values. 
A general functional form which satisfies the above constraints is 

\begin{eqnarray}
\rho(\vecr,\tau;\vec{b})=\rho_0\frac{\tau_0}{\tau} \frac{N_{\textrm{wn}}( t^f_A(\vecr + \vec{b}/2) ,t^f_A(\vecr - \vec{b}/2)   )}{N_\textrm{wn}( t^f_A(0) ,t^f_A(0) )},  \label{wn}
\end{eqnarray}
\nt
where, $N_\textrm{wn}$ is the wounded nucleon participant density \cite{Kolb:2003dz,Nonaka:2006yn}, and $t^f_A(\vecr \pm \vec{b}/2)$
is an input thickness function that may be varied to obtain different spatial profiles $\rho(\vecr,\vec{b},\tau)$. 
The superscript $f$ on the thickness functions is meant to differentiate it from the thickness functions used to 
generate the number of binary collisions in the initial state. 
The time dependence of the density for the first set of computations will be 
assumed to be generated by a pure Bjorken expansion (which leads to a $1/\tau$ dependence in Eq.~\eqref{wn}). 
It may be immediately verified that even within the apparently constraining 
form of Eq.~\eqref{wn}, radically different space-time profiles may be obtained by using different input thickness 
functions.  
The profiles are normalized by the maximum density 
achieved at $\vecr = 0$ in the most central collision at $\vec{b} = 0$ (the denominator of Eq.~\eqref{wn}). 
In such a form, the initial spatial modulation is contained entirely in the normalized modulation 
factor $M(x,y)$ \tie, $\rho(x,y) = \rho_0\frac{\tau_0}{\tau}M(x,y)$. 
By definition, the maximum value of $M(x,y) = 1$.
The unknown factor, $\rho_0 \tau_0$ in the above equation is set in all cases by assuming 
that the $R_{AA}$ for 
the most central collisions at a pion $p_T=8$ GeV be $0.2$. 
Once so normalized, the variation of the $R_{AA}$ with 
$p_T$ for all the different profiles considered for the most central collision is remarkably similar. 
The above model for the density may then be used to predict the $R_{AA }$ at a lesser centrality and also the 
variation of the $R_{AA}$ with the reaction plane.

\begin{figure}[!htbp]
\resizebox{2.5in}{2.5in}{\includegraphics[0.5in,0.5in][5in,5in]{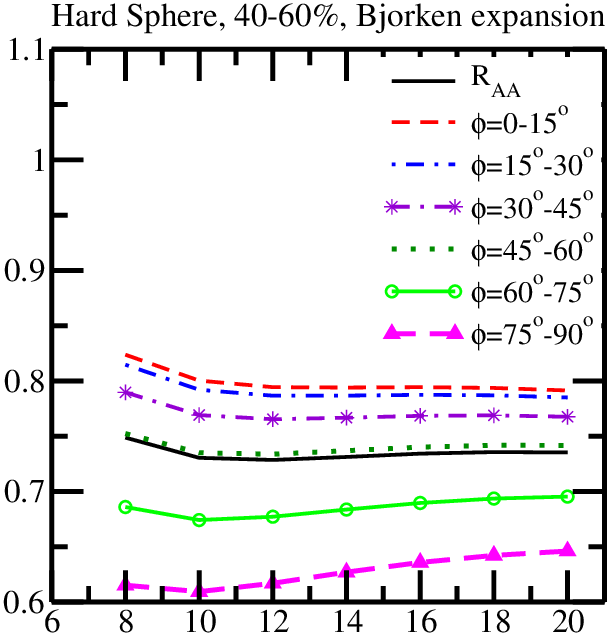}}
\caption{(Color online) $R_{AA}$ at 40-60\% centrality (solid black line) and the variation 
of $R_{AA}$ as a function of the angle $\phi$  with respect to the reaction plane, assuming the 
initial gluon density generated by the overlap of two hard-sphere geometries. The evolution of the 
density is assumed to be purely due to Bjorken expansion. }
\label{hs_den}
\end{figure}

In an effort to demonstrate the sensitivity of the nuclear modification factor to the space-time profile, 
two rather disparate input thickness functions $t^f_A$ are used: the hard sphere and Gaussian. 
The hard sphere thickness function is given as, 

\begin{eqnarray}
t_A^f(\vecr, \vec{b}/2) = C_\textrm{HS} \sqrt{ R_A^2 - (x\pm b_x/2)^2 - (y \pm b_y/2)^2}. \label{t_hs}
\end{eqnarray}
\nt
The focus in this letter will be on the peripheral 
events with a centrality between $40\%-60\%$, there exists a considerable azimuthal asymmetry in the
gluon distribution in such an event for a hard sphere distribution.   
The results of such a calculation are shown in Fig.~\ref{hs_den}. 
In this figure, the solid black line represents the $R_{AA}$ integrated over all emission angles. The 
other lines represent the differential $R_{AA}$ as a function of the angle of the jet with respect to the
reaction plane ($\phi$) \tie, $R_{AA}(\D \phi)$. Due to the large azimuthal asymmetry in the initial 
gluon density, leading to jets
facing different amounts of matter depending on emission angle, there exists a large 
asymmetry as well in the $R_{AA}(\D \phi)$.  The plain dashed line 
$(\phi=0-15^o)$ represents emission in the reaction plane. These represent jets which 
pass through the least amount of matter and as a result exhibit the lowest amount of  modification. 
The dashed line with triangles $(\phi=75-90^o)$ represents emission perpendicular to the plane.
These pass through the maximum amount of matter and thus show the most modification. The 
lines in between represent emission at intermediate angles. This figure clearly demonstrates 
the sensitivity of the nuclear modification factor versus the reaction plane to the azimuthal 
asymmetry of the gluon distribution. 

\begin{figure}[!htbp]
\resizebox{2.5in}{2.5in}{\includegraphics[0.5in,0.5in][5in,5in]{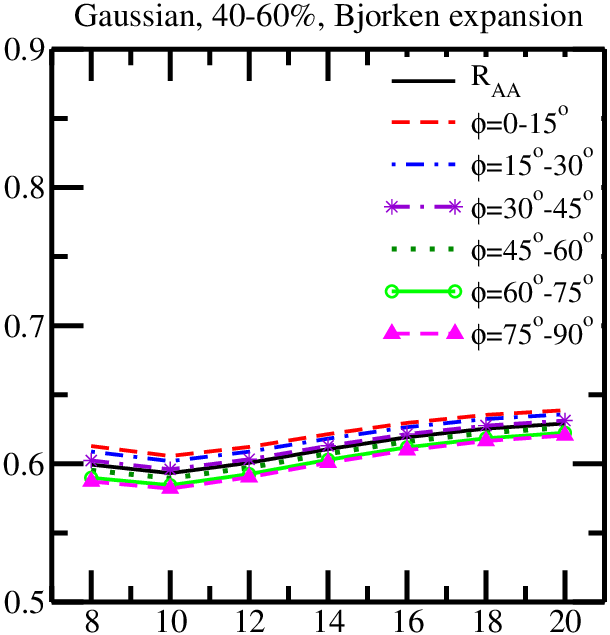}}
\caption{(Color online) Same as Fig.~\ref{hs_den}, with the hard sphere density 
replaced with a Gaussian profile. }
\label{gs_den}
\end{figure}

To put the above statement on a firm footing, the sensitivity of the $R_{AA}(\D \phi)$ is 
tested against a gluon distribution which does not posses as large an azimuthal 
asymmetry as in the case above. This is achieved by using a Gaussian density as the input thickness function, 

\bea
t^f_A(\vecr,\vec{b}) = C_\textrm{G} e^{-\frac{\left| \vecr \pm \vec{b}/2 \right|^2}{2R_G^2}},
\eea
where $R_G$ is an appropriately chosen Gaussian radius. 
As would be expected, such an initial gluon density leads to a very different variation of the 
$R_{AA}$ with respect to the reaction plane. This is plotted in Fig.~\ref{gs_den}. The variation 
of the nuclear modification factor with $\phi$ is much reduced as compared to that from a hard 
sphere thickness function. This difference testifies to the effectiveness of observables 
such as the $R_{AA}(\D \phi)$ as probes of the initial gluon density profile. 
For the initial number of binary collisions, the use of a  hard sphere density is continued. 
This ensures that all jets are produced in the same, finite, well defined, volume.  

The variation of the $R_{AA}$ with $\phi$ may also be sensitive to the temporal evolution of the plasma. 
In both of the preceding calculations, 
the temporal dependence was taken to be that due to Bjorken expansion. In real 
heavy-ion collisions, there is a considerable amount of transverse (radial and elliptic) flow 
of the matter caused by the large pressures built up in the dense matter. To have a completely 
realistic description of such a scenario requires that the density modulation factor $M(x,y,\tau)$ 
be extracted from full three dimensional hydrodynamical simulations \cite{maj_next}.  
In this effort, a simple scaling form will be employed: the two transverse coordinates 
are replaced by $x/[r( \tau) \e(\tau)]$ and $y\e(\tau)/r( \tau)$, where $r(\tau)$ describes the  radial 
expansion and $\e(\tau)$ generates the azimuthally asymmetric expansion. It should be 
noted that there already exists an asymmetry in the density distribution from Eq.~\eqref{wn} where 
the hard sphere thickness function (Eq.~\ref{t_hs}) is used as input to calculate the gluon density.
The factors $r,\e$ are meant to introduce a further time dependent modulation. Thus at $\tau=0$, 
both these factors must be set to unity.
 A simple form for the time dependence of $r$ and $\e$ may be  obtained in a toy model: 
\begin{eqnarray}
r(\tau) = 1 + \frac{v}{R_A}\tau \hspace{0.8cm} \mbox{and} \hspace{0.8cm} \e(\tau)^2 = 1 + \frac{v_\e}{2R_A}\tau.
\end{eqnarray}

\begin{figure}[!htbp]
\resizebox{2.5in}{2.5in}{\includegraphics[0.5in,0.5in][5in,5in]{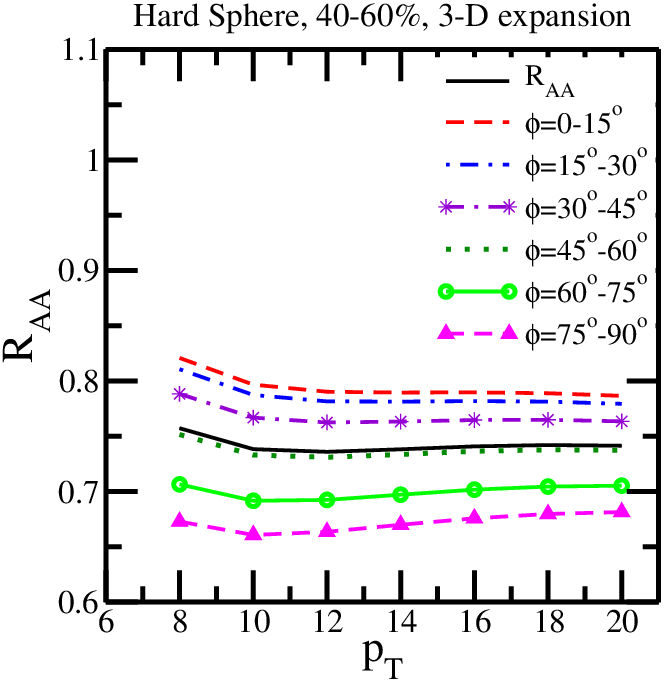}}
\caption{(Color online) Same as Fig.~\ref{hs_den} including transverse flow. }
\label{hs_flow}
\end{figure}

\begin{figure}[!htbp]
\resizebox{2.5in}{2.5in}{\includegraphics[0.5in,0.5in][5in,5in]{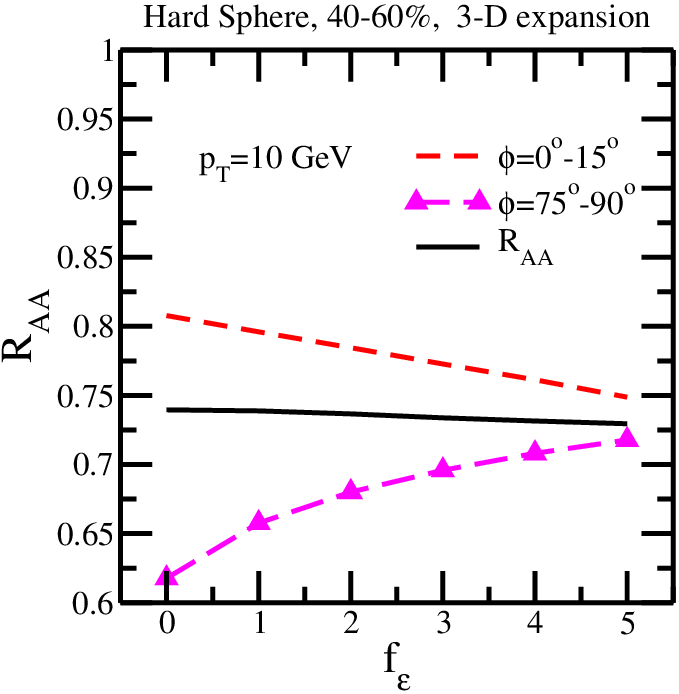}}
\caption{(Color online) Effect of asymmetric flow on the spread of $R_{AA}$ as a function of 
the reaction plane at pion $p_T=10$ GeV. The solid line is the angle integrated $R_{AA}$ the dashed lines 
with (without) triangles  is the $R_{AA}$ with the jet emission angle out of  (in) the reaction plane.}
\label{v2}
\end{figure}

In the above, the radial flow velocity $v$ is taken to be $0.4$ from Ref.~\cite{Kolb:2003dz}. 
The impact parameter dependent asymmetric velocity $v_\e$ is taken to be such as to restore the azimuthal 
symmetry to the boundary of the expanding plasma by $\tau =2 R_A$ \tie, $v_\e = y_{max}(b)/x_{max}(b) -1$.
The maximum $x$ and $y$ coordinate of the overlap at $\tau=0$ is represented as $x_{max},y_{max}$ (note that the use 
of the hard sphere thickness function as input allows a clear identification of an $x_{max}$ and $y_{max}$). 
The integral of the density over the three volume is kept fixed with the replacement 
\begin{eqnarray}
 \rho(x,y,\tau) \ra  \frac{\rho(x,y,\tau)} { r(\tau)^2 }.
\end{eqnarray}
\nt
The  result of such a model calculation is shown in Fig.~\ref{hs_flow}. 
As is clearly demonstrated, the large asymmetric flow, leads to an $R_{AA}$ 
versus reaction plane with a lesser spread as a function of $\phi$ than in Fig.~\ref{hs_den}. 
The ordering of the contributions remains unchanged. Thus even with a realistic 
amount of radial and elliptic flow, the $R_{AA}(\D \phi)$ still maintains a 
sensitivity to the initial asymmetry. If the asymmetric expansion is increased, 
this leads to a further reduction in the spread. This is shown in Fig.~\ref{v2} where 
the $R_{AA}$,  along with the $R_{AA}$ in reaction plane and out of plane are 
plotted as a function of  $f_\e$, an overall  multiplicative factor used to modify the asymmetric velocity $v_\e$. 
In the figure,  $f_\e = 0$ corresponds 
to the case of Fig.~\ref{hs_den} with a radial flow, \tie, with no elliptic flow. 
The case of $v_\e=1$ corresponds to the case of Fig.~\ref{hs_flow}, \tie, with elliptic flow 
parametrized from Ref.~\cite{Kolb:2003dz}. Higher values of $f_\e$
represent multiples of the asymmetric velocity used in Fig.~\ref{hs_flow}.
As expected, raising the asymmetric velocity leads to a decrease in the 
spread of the $R_{AA}$ as a function of the reaction plane.
It should be pointed out that in both Figs.~\ref{hs_den} and \ref{hs_flow}, 
the final state gluon density profile starts out with a considerable azimuthal asymmetry as a hard 
sphere thickness function is used as input in Eq.~\eqref{wn}. 

The calculations presented above offer sufficient evidence that differential probes such 
as the $R_{AA}$ versus the reaction plane offer deeper insight into the space-time 
profile of the matter produced at RHIC. 
Sharper eccentricities lead to large variations of the $R_{AA}$ with respect to the emission 
angle. The effect of large elliptic flow is discernable. 
It is true that such observations alone will 
not distinguish the effect of a large elliptic flow from that of a lower eccentricity in   
the initial condition. To discern between such facets will require multiple differential 
probes to be compared in unison. The nuclear modification factor versus the reaction 
plane carries more information than the azimuthal asymmetry of high $p_T$ hadrons
(the  $v_2$ at high $p_T$).  
The reader will note that the $v_2$ obtained from the $R_{AA}(\D \phi)$
(Figs.~\ref{hs_den}-\ref{hs_flow}) is insensitive to the overall magnitude of the 
$R_{AA}$.
%  \tie, $R_{AA}(\D \phi)$ and $n R_{AA} (\D \phi)$ produce the same $v_2$, where 
% $n$ is an overall constant.

The current exercise is meant to serve as 
a first attempt in the resolution of the spatio-temporal profile of the hot plasma. 
As such, a variety of approximations were made, $k_T$ broadening was ignored, and 
the number of initial binary collisions was taken from a hard sphere nuclear density distribution. 
This was done in the interest of  focussing the study on the effect of the plasma  profile on the 
variation of $R_{AA}$ with respect to the jet emission angle. The inclusion of such effects 
will result is slender modifications to the results outlined in this letter and are left for a future, 
more rigorous comparison with data.

{\em Acknowledgments:}  This work was supported in part by the U.S. Department of Energy
under grant DE-FG02-05ER41367. The author thanks S.~A.~Bass, B.~M\"{u}ller and X.~N.~Wang 
for enlightening discussions.


\begin{thebibliography}{99}

%\cite{Karsch:2003jg}
\bibitem{Karsch:2003jg}
  F.~Karsch and E.~Laermann,
  %``Thermodynamics and in-medium hadron properties from lattice QCD,''
  arXiv:hep-lat/0305025.
  %%CITATION = HEP-LAT 0305025;%%





\bibitem{RHIC_Whitepapers}
  I.~Arsene {\em et al.}, 
  %Quark-Gluon Plasma and the Color Glass Condensate at RHIC?  
  %The Perspective from the BRAHMS Experiment.
  Nucl.\ Phys.\ A {\bf 757}, 1 (2005);
  %nucl-ex/0410020
  B.~B.~Back {\em et al.},
  %The PHOBOS Perspective on Discoveries at RHIC.
  {\em ibid.} {\bf 757}, 28 (2005);
  %nucl-ex/0410022
  J.~Adams {\em et al.},
  %Experimental and Theoretical Challenges in the Search for the Quark 
  %Gluon Plasma:  The STAR Collaboration's Critical Assessment of the 
  %Evidence from RHIC Collisions.
  {\em ibid.} {\bf 757}, 102 (2005);
  K.~Adcox {\em et al.},
  %Formation of Dense Partonic Matter in Relativistic Nucleus-Nucleus 
  %Collisions at RHIC:
  %Experimental Evaluation by the PHENIX Collaboration.
  {\em ibid.} {\bf 757}, 184 (2005).
  %nucl-ex/0410003


%\cite{Braun-Munzinger:2003zd}
\bibitem{Braun-Munzinger:2003zd}
  P.~Braun-Munzinger, K.~Redlich and J.~Stachel,
  %``Particle production in heavy ion collisions,''
  arXiv:nucl-th/0304013 (and references therein); 
  %%CITATION = NUCL-TH 0304013;%%

%\cite{Kolb:2003dz}
\bibitem{Kolb:2003dz}
  P.~F.~Kolb and U.~W.~Heinz,
  %``Hydrodynamic description of ultrarelativistic heavy-ion collisions,''
  arXiv:nucl-th/0305084.
  %%CITATION = NUCL-TH 0305084;%%

%\cite{Huovinen:2003fa}
\bibitem{Huovinen:2003fa}
  P.~Huovinen,
  %``Hydrodynamical description of collective flow,''
  arXiv:nucl-th/0305064.
  %%CITATION = NUCL-TH 0305064;%%
%\cite{Gyulassy:2003mc}


%\cite{Teaney:2003kp}
\bibitem{Teaney:2003kp}
  D.~Teaney,
  %``Effect of shear viscosity on spectra, elliptic flow, and Hanbury
  %Brown-Twiss radii,''
  Phys.\ Rev.\ C {\bf 68}, 034913 (2003)
  [arXiv:nucl-th/0301099].
  %%CITATION = NUCL-TH 0301099;%%




\bibitem{Gyulassy:2003mc}
M.~Gyulassy, I.~Vitev, X.~N.~Wang and B.~W.~Zhang,
%``Jet quenching and radiative energy loss in dense nuclear matter,''
arXiv:nucl-th/0302077;
%%CITATION = NUCL-TH 0302077;%%
X.~N.~Wang,
%``Discovery of jet quenching and beyond,''
arXiv:nucl-th/0405017;
%%CITATION = NUCL-TH 0405017;%%
%\cite{Baier:1996kr}
% \bibitem{Baier:1996kr}
  R.~Baier, Y.~L.~Dokshitzer, A.~H.~Mueller, S.~Peigne and D.~Schiff,
  %``Radiative energy loss of high energy quarks and gluons in a  finite-volume
  %quark-gluon plasma,''
  Nucl.\ Phys.\ B {\bf 483}, 291 (1997)
  [arXiv:hep-ph/9607355];
  %%CITATION = HEP-PH 9607355;%%
%\cite{Baier:1996sk}
%\cite{Wiedemann:2000za}
% \bibitem{Wiedemann:2000za}
  U.~A.~Wiedemann,
  %``Gluon radiation off hard quarks in a nuclear environment: Opacity
  %expansion,''
  Nucl.\ Phys.\ B {\bf 588}, 303 (2000)
  [arXiv:hep-ph/0005129];
  %%CITATION = HEP-PH 0005129;%%
%\cite{Turbide:2005fk}
% \bibitem{Turbide:2005fk}
  S.~Turbide, C.~Gale, S.~Jeon and G.~D.~Moore,
%    ``Energy loss of leading hadrons and direct photon production in evolving
  %quark-gluon plasma,''
  Phys.\ Rev.\ C {\bf 72}, 014906 (2005)
  [arXiv:hep-ph/0502248].
  %%CITATION = HEP-PH 0502248;%%



%\cite{Gyulassy:2004zy}
\bibitem{Gyulassy:2004zy}
  M.~Gyulassy and L.~McLerran,
  %``New forms of QCD matter discovered at RHIC,''
  Nucl.\ Phys.\ A {\bf 750}, 30 (2005)
  [arXiv:nucl-th/0405013].
  %%CITATION = NUCL-TH 0405013;%%



%\cite{Nonaka:2006yn}
\bibitem{Nonaka:2006yn}
  C.~Nonaka and S.~A.~Bass,
  %``Space-time evolution of bulk QCD matter,''
  arXiv:nucl-th/0607018.
  %%CITATION = NUCL-TH 0607018;%%

%\cite{Armesto:2004pt}
\bibitem{Armesto:2004pt}
  N.~Armesto, C.~A.~Salgado and U.~A.~Wiedemann,
  %``Measuring the collective flow with jets,''
  Phys.\ Rev.\ Lett.\  {\bf 93}, 242301 (2004)
  [arXiv:hep-ph/0405301].
  %%CITATION = HEP-PH 0405301;


\bibitem{highpt}
K.~Adcox {\it et al.}  [PHENIX Collaboration],
 %``Suppression of hadrons with large transverse momentum in central  Au + Au
%collisions at s**(1/2)(N N) = 130-GeV,''
Phys.\ Rev.\ Lett.\  {\bf 88}, 022301 (2002)
[arXiv:nucl-ex/0109003];
%%CITATION = NUCL-EX 0109003;%%
C.~Adler {\it et al.} [STAR Collaboration],
 %``Centrality dependence of high p(T) hadron suppression in Au + Au collisions
%at s(NN)**(1/2) = 130-GeV,''
Phys.\ Rev.\ Lett.\  {\bf 89}, 202301 (2002)
[arXiv:nucl-ex/0206011].
%%CITATION = NUCL-EX 0206011;%%

%\cite{Gyulassy:2000fs}
\bibitem{Gyulassy:2000fs}
  M.~Gyulassy, P.~Levai and I.~Vitev,
  %``Non-Abelian energy loss at finite opacity,''
  Phys.\ Rev.\ Lett.\  {\bf 85}, 5535 (2000)
  [arXiv:nucl-th/0005032];
  %%CITATION = NUCL-TH 0005032;%%




%\cite{Adams:2006yt}
\bibitem{Adams:2006yt}
  J.~Adams {\it et al.}  [STAR Collaboration],
%    ``Direct observation of dijets in central Au + Au collisions at s(NN)**(1/2)
  %= 200-GeV,''
  arXiv:nucl-ex/0604018;
  %%CITATION = NUCL-EX 0604018;%%
%\cite{Adler:2005ee}
% \bibitem{Adler:2005ee}
  S.~S.~Adler {\it et al.}  [PHENIX Collaboration],
%    ``Modifications to di-jet hadron pair correlations in Au + Au collisions  at
  %s(NN)**(1/2) = 200-GeV,''
  arXiv:nucl-ex/0507004.
  %%CITATION = NUCL-EX 0507004;%%

%\cite{Vitev:2005yg}
\bibitem{Vitev:2005yg}
  I.~Vitev,
  %``Large angle hadron correlations from medium-induced gluon radiation,''
  Phys.\ Lett.\ B {\bf 630}, 78 (2005)
  [arXiv:hep-ph/0501255];
  %%CITATION = HEP-PH 0501255;%%
%\cite{Casalderrey-Solana:2004qm}
% \bibitem{Casalderrey-Solana:2004qm}
  J.~Casalderrey-Solana, E.~V.~Shuryak and D.~Teaney,
  %``Conical flow induced by quenched QCD jets,''
  J.\ Phys.\ Conf.\ Ser.\  {\bf 27}, 22 (2005)
  [arXiv:hep-ph/0411315];
  %%CITATION = HEP-PH 0411315;%%%
%\cite{Ruppert:2005uz}
% \bibitem{Ruppert:2005uz}
  J.~Ruppert and B.~Muller,
  %``Waking the colored plasma,''
  Phys.\ Lett.\ B {\bf 618}, 123 (2005)
  [arXiv:hep-ph/0503158];
  %%CITATION = HEP-PH 0503158;%%
%\cite{Majumder:2004wh}
% \bibitem{Majumder:2004wh}
  A.~Majumder and X.~N.~Wang,
  %``The dihadron fragmentation function and its evolution,''
  Phys.\ Rev.\ D {\bf 70}, 014007 (2004)
  [arXiv:hep-ph/0402245];
  %%CITATION = HEP-PH 0402245;%%
%\cite{Majumder:2004br}
% \bibitem{Majumder:2004br}
  A.~Majumder and X.~N.~Wang,
  %``Evolution of the parton dihadron fragmentation functions,''
  Phys.\ Rev.\ D {\bf 72}, 034007 (2005)
  [arXiv:hep-ph/0411174];
  %%CITATION = HEP-PH 0411174;%%
%\cite{Koch:2005sx}
% \bibitem{Koch:2005sx}
  V.~Koch, A.~Majumder and X.~N.~Wang,
  %``Cherenkov radiation from jets in heavy-ion collisions,''
  Phys.\ Rev.\ Lett.\  {\bf 96}, 172302 (2006)
  [arXiv:nucl-th/0507063];
  %%CITATION = NUCL-TH 0507063;%%
%\cite{Majumder:2005sw}
%\bibitem{Majumder:2005sw}
  A.~Majumder and X.~N.~Wang,
  %``LPM interference and Cherenkov-like gluon bremsstrahlung in dense matter,''
  Phys.\ Rev.\ C {\bf 73}, 051901 (2006)
  [arXiv:nucl-th/0507062].
  %%CITATION = NUCL-TH 0507062;%%


%\cite{Renk:2006qg}
\bibitem{Renk:2006qg}
  T.~Renk,
  %``Towards jet tomography: gamma hadron correlations,''
  arXiv:hep-ph/0607166.
  %%CITATION = HEP-PH 0607166;%%

%\cite{Collins:1989gx}
\bibitem{Collins:1989gx}
  J.~C.~Collins, D.~E.~Soper and G.~Sterman,
  %``Factorization of Hard Processes in QCD,''
  Adv.\ Ser.\ Direct.\ High Energy Phys.\  {\bf 5}, 1 (1988)
  [arXiv:hep-ph/0409313].
  %%CITATION = HEP-PH 0409313


%\cite{Wang:2003aw}
\bibitem{Wang:2003aw}
  X.~N.~Wang,
  %``Why the observed jet quenching at RHIC is due to parton energy loss,''
  Phys.\ Lett.\ B {\bf 579}, 299 (2004)
  [arXiv:nucl-th/0307036].
  %%CITATION = NUCL-TH 0307036;%%

%\cite{Li:2001xa}
\bibitem{Li:2001xa}
  S.~y.~Li and X.~N.~Wang,
  %``Gluon shadowing and hadron production at RHIC,''
  Phys.\ Lett.\ B {\bf 527}, 85 (2002)
  [arXiv:nucl-th/0110075].
  %%CITATION = NUCL-TH 0110075;%%




\bibitem{guowang} X.~F.~Guo and X.~N.~Wang,
 %``Multiple scattering, parton energy loss and modified fragmentation
%functions in deeply inelastic e A scattering,''
Phys.\ Rev.\ Lett.\  {\bf 85}, 3591 (2000)
[arXiv:hep-ph/0005044];
%%CITATION = HEP-PH 0005044;%%
X.~N.~Wang and X.~F.~Guo,
%``Multiple parton scattering in nuclei: Parton energy loss,''
Nucl.\ Phys.\ A {\bf 696}, 788 (2001)
[arXiv:hep-ph/0102230];
%%CITATION = HEP-PH 0102230;%%
B.~W.~Zhang and X.-N.~Wang,
%``Multiple parton scattering in nuclei: Beyond helicity amplitude
%approximation,''
Nucl.\ Phys.\ A {\bf 720}, 429 (2003)
[arXiv:hep-ph/0301195].
%%CITATION = HEP-PH 0301195;%%

\bibitem{bin95} 
J.~Binnewies, B.~A.~Kniehl and G.~Kramer,
%``Pion and kaon production in e+ e- and e p collisions at next-to-leading
%order,''
Phys.\ Rev.\ D {\bf 52}, 4947 (1995)
[arXiv:hep-ph/9503464].
%%CITATION = HEP-PH 9503464;%%


\bibitem{lqs}
%\cite{Luo:1992fz}
M.~Luo, J.~w.~Qiu and G.~Sterman,
%``Nuclear dependence at large transverse momentum,''
Phys.\ Lett.\ B {\bf 279}, 377 (1992);
%%CITATION = PHLTA,B279,377;%%
%\cite{Luo:1994np}
%\bibitem{Luo:1994np}
M.~Luo, J.~w.~Qiu and G.~Sterman,
%``Anomalous nuclear enhancement in deeply inelastic scattering and
%photoproduction,''
Phys.\ Rev.\ D {\bf 50}, 1951 (1994).
%%CITATION = PHRVA,D50,1951;%%





%\cite{Majumder:2004pt}
\bibitem{maj04e}
A.~Majumder, E.~Wang and X.~N.~Wang,
%``Modified dihadron fragmentation functions in hot and nuclear matter,''
arXiv:nucl-th/0412061;
%%CITATION = NUCL-TH 0412061;%%
%\cite{Majumder:2005vs}
% \bibitem{Majumder:2005vs}
  A.~Majumder,
  %``Jet quenching: The medium modification of the single and double
  %fragmentation functions,''
  arXiv:nucl-th/0501029.
  %%CITATION = NUCL-TH 0501029;%%



%\cite{Baier:1996sk}
\bibitem{Baier:1996sk}
  R.~Baier, Y.~L.~Dokshitzer, A.~H.~Mueller, S.~Peigne and D.~Schiff,
   %``Radiative energy loss and p(T)-broadening of high energy partons in
  %nuclei,''
  Nucl.\ Phys.\ B {\bf 484}, 265 (1997)
  [arXiv:hep-ph/9608322].
  %%CITATION = HEP-PH 9608322;%%



\bibitem{maj_next}
A.~Majumder, C.~Nonaka and S.~Bass, \emph{in preparation}.


\end{thebibliography}
\end{document}